\documentclass[12pt]{iopart}
\usepackage{graphicx}
\newcommand{\figwid}{9cm}
\newcommand{\tfigwid}{7cm}
\newcommand{\lfigwid}{13cm}

\begin{document}

\title[Charge and spin dynamics in a 2DEG]{Charge and spin dynamics in a two dimensional electron gas}

\author{A Pug\v{z}lys$^1$\footnote{present address:Photonics Institute, Vienna University
of Technology, Gusshausstrasse 27/387,
1040 Vienna, Austria}, P J Rizo$^1$, K Ivanin$^1$, A Slachter$^1$, D Reuter$^2$,\\
A D Wieck$^2$, C H van der Wal$^1$ and P H M van Loosdrecht$^1$}
\address{$^1$ MSC$^{plus}$, University of Groningen,
Nijenborgh 4, 9747 AG Groningen, The Netherlands}
\address{$^2$ Angewandte Festk\"{o}rperphysik, Ruhr-Universit\"{a}t Bochum, D-44780 Bochum, Germany}
\ead{p.h.m.van.loosdrecht@rug.nl}

\begin{abstract}
A number of time resolved optical experiments probing and controlling the spin
and charge dynamics of the high mobility two-dimensional electron gas in a
GaAs/AlGaAs heterojunction are discussed. These include time resolved
reflectivity, luminescence, transient grating, magneto-optical Kerr effect,
and electro-optical Kerr effect experiments. The optical
experiments provide information on the carrier lifetimes and spin dephasing
times, as well as on the carrier diffusion coefficient which directly gives the
charge mobility. A combination of the two types of Kerr experiments,
shows to be useful in extracting both the carrier lifetimes and spin dephasing
times in a single experiment.
\end{abstract}
\pacs{73.40.Kp, 42.50.Md,	78.47.+p, 73.43.Fj, 78.66.Fd}

\section{Introduction\label{sec:intro} }

Realization of the dream of spintronics \cite{Wolf01,Zutic04} requires not only
detailed knowledge on how to inject electron spin currents into functional
devices \cite{Schmidt05}, but also profound understanding of how the spin
information eventually gets lost, how it is connected to the charge degree of
freedom, and how to manipulate and control the spin degree of freedom. Some
useful spin-based devices are available, but the final grail of  fully
spin-based electronics is still not reached, despite tremendous progress in the
field \cite{Zutic04}. Electronic and optical methods are the two favourite
manners to study and manipulate the spin degree of freedom in bulk and confined
semiconductors. In particular optical methods are quite powerful, since they
can address {\em and} manipulate the charge and spin degrees of freedom
independently. Quite early on it has been shown using transient Faraday
spectroscopy that the lifetime of a coherent spin population in $n$-doped bulk
GaAs can be tremendously long, exceeding 100 nanoseconds \cite{Kikkawa98}.
Moreover, it has been shown that such a coherent spin population can be
transported over tens of micrometers before eventually the coherence is lost
\cite{Kikkawa99,Kikkawa01}. Even though these bulk results are instrumental in
the study of spin coherence and control, it is clear, in the light of
functional devices, that the properties of confined structures as found in
quantum well systems, heterojunctions, and quantum wires and dots are even more
important.\cite{Kikkawa97,Holleitner06,Hanson03} 
Again, optical methods provide excellent tools to study the charge
and spin dynamics of these confined structures. The charge channel can for
instance be studied using time resolved luminescence, reflectivity, and
transient grating techniques. The spin degrees of freedom can be addressed
using time resolved magneto-optical Kerr effect, Faraday, and
transient spin grating techniques. Finally using a combination of
magneto-optical and birefringence methods, one can even address both the charge
and spin degrees of freedom simultaneously.

This paper presents and compares results from some of these optical techniques
as applied to a high mobility two-dimensional electron gas in a GaAs/AlGaAs
heterojunction (HJ2DEG). The different techniques were applied with the same
laser system and sample material, allowing a qualitative and quantitative
comparison of the results from the different techniques.
Moreover, part of the experiments have been
performed under identical conditions on both the 2DEG material and bulk GaAs
material. This demonstrates how the techniques are useful for
studies on 2DEG systems, and evidences that the signals are from the 2DEG
system rather than from the surrounding bulk material. Developing a
consistent picture of the coupled spin and charge dynamics in the GaAs/AlGaAs
multilayer system using only one method is very challenging. 
Therefore it is better to use results from different methods to collect 
evidence for the mechanisms that underlie this spin and charge dynamics.

The advantage of using heterojunctions for charge transport is obviously their
high mobility, which can be 10's of millions cm$^{2}$/Vs at low
temperatures. For spin transport, this advantage turns out to be a
disadvantage. The long mean free path results in a dephasing of a macroscopic
coherent spin state trough the coupling to the anisotropic crystal field,
leading to dephasing times of the order of a nanosecond or less. Note however,
that optical measurements on heterojunction 2DEGs are challenging because the
2DEG has transition energies in the same spectral region as the underlying
buffer layer. This makes optical studies in HJ2DEGs more involved than those
on double sided quantum wells (QWs), which can have transition energies that
differ substantially from all other characteristic transitions in the QW structure. As a
result, it is for QWs relatively easy to spectrally discern the optical
response originating from the QW. For heterojunction 2DEGs however, one has to
disentangle the bulk and 2DEG contributions from the mixed 2DEG/bulk optical
response, and this will therefore be discussed for several of the optical
techniques used in the experiments.

After an introduction of the samples and the experimental set-up in 
section~\ref{sec:exp}, the charge and spin dynamics of the HJ2DEG will be discussed in
sections~\ref{sec:charge} and \ref{sec:spin}, respectively. 
Next, in section~\ref{sec:spincharge}, the electro-optical Kerr effect in the presence of an
external magnetic field (the MEOKE technique) applied to HJ2DEGs will be
discussed. This technique allows for the simultaneous study of both the charge and spin
dynamics. Finally, section~\ref{sec:concl} will summarize and conclude this
paper.

\section{Sample material and experimental set-up\label{sec:exp}}

\subsection{GaAs/AlGaAs Heterojunction sample}

\begin{figure}[htb]
\begin{center}
\includegraphics[width=\figwid]{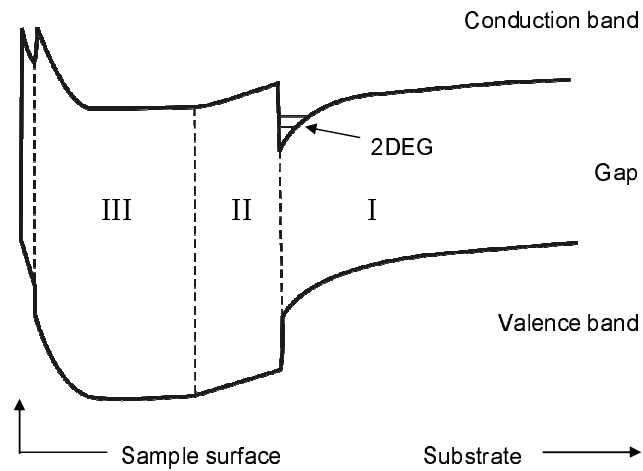}
\end{center}
\caption{Schematic of the conduction and valence bands along the growth
direction (here from right to left) of a heterostructure containing a
heterojunction 2DEG. The roman numerals I, II, and III indicate the buffer,
spacer, and dopant layers, respectively. The illustration also shows the capping
layer above the dopant layer. The 2DEG is localized at the notch potential
between the buffer and spacer layers. Notice that a built-in electric field is
established across the buffer layer.} \label{bandstructure}
\end{figure}

Investigations of the intriguing properties of two-dimensional electron gases
(2DEGs) such as the integer and fractional quantum Hall effect have stimulated
extensive optical studies of modulation doped single heterojunctions (see
\cite{Kukushkin96} and references therein). 
These heterojunctions show
remarkably high carrier mobilities due to the separation of the free carriers
from the parent ionized donors and the ability to grow AlGaAs on top of GaAs
with extremely low interface roughness\cite{Stormer79}. Heterojunction 2DEGs
are formed by growing an undoped narrow band gap semiconductor, known as the
buffer (or active) layer, in contact with a doped wide band gap semiconductor
(or dopant) layer. An undoped spacer layer made of the wide band gap material
deposited between the dopant and buffer layers enhances carrier mobility by
reducing Coulomb scattering between free carriers and ionized donors. Excess
carriers from the dopant layer reduce their energy by a transfer to the
conduction band of the narrow band gap semiconductor. Here, these free carriers
accumulate at the heterointerface due to electrostatic attraction from the
parent ionized donors in the dopant layer, thus forming a 2DEG 
(see figure~\ref{bandstructure}). 
The resulting charge separation establishes an electric
field or band bending along the growth direction in the buffer layer. The
preferred materials for high mobility 2DEGs are GaAs (buffer layer) and an
Al$_x$Ga$_{1-x}$As alloy (dopant and spacer layers). The reasons for utilizing
these materials are the matching lattice constants, the good interface quality,
and the low effective electron mass in the conduction band of GaAs.
Note that the term heterojunction will be used for the GaAs/AlGaAs interface where the 2DEG is confined, 
while the term heterostructure will be used for the entire GaAs/AlGaAs multilayer
system.

The experiments described in this paper are performed on a modulation doped
single heterojunction with a band structure as depicted in 
figure~\ref{bandstructure}. The heterostructure consists of 9330~\AA\ undoped GaAs
buffer layer grown on top of a (100) $i$-GaAs substrate. On top of this, 368~\AA\
of undoped Al$_{.32}$Ga$_{.68}$As forms the spacer layer. The dopant layer
consists of 719~\AA\ of Si-doped Al$_{.32}$Ga$_{.68}$As with $3\times10^{18}$
dopants/cm$^{3}$. The heterostructure is capped with 55~\AA\ of $n$-GaAs. The
dark mobility and 2DEG carrier density, as derived from transport experiments at 4.2 K are
$1.59\times 10^{6}$~cm$^{2}$/Vs and $2.14\times 10^{11}$~cm$^{-2}$ respectively.
After illumination these values become
$2.7\times 10^{6}$~cm$^{2}$/Vs and $4.2\times 10^{11}$~cm$^{-2}$.

\subsection{Laser system and optical cryostat}

In the experiments a cavity dumped Titanium sapphire laser (Cascade, Kapteyn-Murnane
Laboratories Inc.) was used as a source of excitation and probe light. The
laser generates 25~fs, 40~nJ pulses centred around 780~nm. The repetition rate
of the laser pulses can be tuned between 40~kHz and 4~MHz. Experiments were
performed either by exciting and probing the sample with a broad spectrum
corresponding to a 25~fs pulse or with spectral portions having width of
$\sim$10 nm (90~\% transmission), 
which were selected using interference filters at appropriate places
in the optical set up. In the latter case, the temporal width was about 120~fs.
The typical focal spot size in the experiments is about 
$\sim$75~$\mu$m in diameter, unless stated differently.
The energy of the
excitation pulses was varied between 100~pJ and 1.2~nJ. Experiments were
carried out at temperatures ranging from room temperature down to 4.2 K. In the case of
low temperature measurements the sample was placed in a optical cryostat with a
split pair superconducting magnet system (Spectromag, Oxford Instruments), which allows to
vary sample temperature from 2 K up to 300 K and is capable of creating an
external magnetic field up to 8 Tesla.

\section{Charge dynamics\label{sec:charge}}

\subsection{Photoluminescence\label{sec:pl}}

Photoluminescence (PL) studies are very well suited for identifying charge 
relaxation and recombination mechanisms and their time scales. 
However, it is well known that PL signals from HJ2DEG samples are dominated by signals 
from the underlying bulk regions, and it is very difficult to extract the weak 
signals that are due the 2DEG system.
Time- and spectrally-resolved PL studies on a HJ2DEG
sample, mounted in a cold finger cryostat, 
have been performed using streak-camera detection . 
The results (no data presented here) show indeed strong PL signals that must be
attributed to processes in the bulk substrate, and only a very weak
signal from the HJ2DEG at high pump intensities.  
Luminescence from the bulk GaAs buffer layer can be affected by
the presence of the 2DEG and the built-in electric field in the
buffer layer.\cite{shen99} Time-resolved PL measurements in the 2DEG sample
revealed several PL lines from the GaAs buffer layer.  
The lifetime of the brightest PL
line, the bound exciton line,is strongly dependent on
temperature and showed a very slow growth at early times. At 4.2 K
after illumination with 780 nm, 120 fs pulses the PL intensity
reaches a maximum after 2.3 ns. The PL lifetime measured was 2.35
ns. At 40 K, the peak intensity delay and PL lifetime becomes 
0.9 ns and 9 ns, respectively. For comparison, an i-GaAs sample was
measured under identical excitation conditions and the parameters
of the bound exciton line were extracted. In this case the peak
intensity delay and PL lifetime at 4.2 K were 0.3 ns and 0.36 ns
respectively. At 40 K the decay time remained unchanged within
experimental accuracy. 
Further, the results are
consistent with published results from earlier work (discussed below). Since
this type of results will be of interest for the discussion of the results from other 
methods, a summary of the main findings reported in literature  
will be given here. Note however, that while much results have been published on this
topic, the exact process responsible for the PL associated with 2D electrons in
samples with an undoped buffer layer still remains unclear.

In HJ2DEGs the carrier accumulation at the heterointerface bends the buffer
layer conduction band below the Fermi level in the region close to the
heterojunction \cite{Stormer79}, as depicted in figure~\ref{bandstructure}. This
built in electric field rapidly segregates photoexcited electrons toward the
heterointerface and holes toward the back of the buffer layer. 2D electron and
3D hole recombination in these heterostructures is then an indirect process in
real space. Spatially indirect recombination coupled with the space charge
potential present in the buffer layer give rise to a red shift in the PL energy
akin to the quantum confined Stark effect \cite{Miller84}. In view of the rapid
segregation of electrons and holes in heterojunction 2DEGs it is surprising
that any holes are present at the heterointerface to recombine with 2D
electrons. Significant efforts have been made to explain the presence of holes
near the heterojunction, but the mechanism by which segregated electrons and holes
recombine in a heterojunction 2DEG still remains controversial.

Early PL studies identified the bands particular to heterojunction structures
with a 2DEG and described their features. Yuan \textit{et al.} studied the PL
from heterojunctions with different layer thickness and Al-composition in the
barriers \cite{Yuan85}. They observed a PL band associated with the
heterojunction which they called the H-band, with peak energies between $\sim$1.505 eV and $\sim$1.525 eV (at 1.4 K), depending on excitation density and the Al-composition of the barriers. The
shift of the PL peak energy with excitation density is a trait intimately
linked to the shape of the potential well in heterojunctions. Several other authors have
reported a similar PL peak energy shift with excitation density (see for
example \cite{Zhao90} and \cite{Kim95}). In heterojunction 2DEGs, segregated
photoexcited carriers screen the built-in electric field thus flattening the
buffer layer bands. 2D electron-3D hole recombination is still indirect in real
space but a reduction of the built-in electric field reduces the red shift 
(produces a blue shift) of the luminescence band. 
Higher excitation densities generate more
photoexcited electrons and holes that segregate and oppose the built-in
electric field thereby increasing the H-band's peak energy. Band gap
renormalization due to the high concentration of free carriers at the
heterojunction also contributes to the observed spectral shifts.

A blue shift in the H-band PL is also observed when the built-in electric field
in the buffer layer is lowered by applying an external bias via a top gate\cite{Zhao90}. 
Reversing the gate bias produces a red shift of the H-band by
increasing the band bending \cite{Zhao91}. Other PL bands, originating from the
buffer layer, show no spectral shifts under intense illumination nor under the
influence of an external electric field \cite{Yuan85, Kim95, Yang88}. The
presence of PL lines characteristic of the bulk buffer material highlights the
fact that in heterojunction structures 2D carriers are in close relation with
bulk carriers. In fact a flat band region can exist in the buffer layer away
from the heterojunction which is actually a layer of bulk material. The
existence of this flat band region depends on temperature, 2DEG carrier
density, buffer layer thickness and unintentional acceptor doping level.

In view of the inevitable interaction of the excitation beam with
the buffer layer it is necessary to establish that the H-band is
indeed a recombination involving 2D electrons. This was
conclusively demonstrated by Kukushkin et al. \cite{Kukushkin88}
through the modifications of the H-band of a 2DEG in the quantum
Hall regime. In their work the authors showed that the PL
intensity at the Fermi energy exhibit the same Shubnikov-de~Haas
oscillations as detected by magneto-transport measurements. By
tilting the magnetic field direction, they determined that the
observed oscillations depended only on the component of the
magnetic field normal to the plane of the heterojunction, a proof of the 2D
character of the carriers involved in the PL. Later, Turberfield
et al. \cite{Turberfield90} and Buhmann et al. \cite{Buhmann90}
observed the quantum Hall effect at integer and fractional filling
factors by modifications to the detected PL from 2D electrons in a
heterojunction.

The 2D carrier concentration in the heterojunction can be altered by
illumination. A persistent increase in 2DEG density (persistent
photoconductivity (PPC)) was observed by St\"{o}rmer \textit{et al.}
\cite{Stormer79} after illuminating the sample with photons of energy below the
AlGaAs band gap. The additional electrons are photoexcited from deep traps
called \textit{DX} centres in the AlGaAs. Some of the electrons released tunnel
into the GaAs and accumulate in the heterojunction potential well. Chaves
\textit{et al.} \cite{Chaves86} and Chou \textit{et al.} \cite{chou85} observed
negative persistent photoconductivity (NPPC) in 2D electron and hole gases in
modulation doped quantum wells respectively. In this case, illumination with
photons above the AlGaAs band gap is necessary. Photoexcited holes in the AlGaAs
are swept into the GaAs by the built in electric field in the spacer layer
without having to cross an energy barrier (see figure~\ref{bandstructure}). The
holes are then trapped in the GaAs close to the heterojunction and recombine
with 2D electrons thus reducing the 2DEG density. NPPC and the magneto-optical
experiments discussed above demonstrate that 2D electrons can recombine with
photoexcited holes even though both carrier types are segregated by the
built-in potential in the buffer layer.

Time resolved PL studies have yielded additional information about photoexcited
carrier dynamics in single heterojunctions. After short pulse excitation photoexcited
carriers drift in opposite directions driven by the built-in electric field. A
photoexcited hole with mobility $\mu_{h}\sim10^{4}$~cm/Vs in a moderately low
built-in electric field $E_{B}$ of $10^{3}$~V/cm will drift a distance $d$ of 1
micron away from the heterojunction in roughly $d/\mu_{h}E_{B}=10$ ps.
Time-resolved PL measurements can observe the dynamics after the segregated
charges have equilibrated. Bergman \textit{et al.} \cite{Bergman91} measured
the time dependence of the H-band PL peak energy. The authors reported a
red shift of the H-band peak energy as a function of time. This should be compared
to the continuous wave measurements as a function of excitation density. Reducing the
excitation density in this case shifts the H-band PL to lower peak energies
\cite{Yuan85, Zhao90, Kim95}. In the time resolved measurements the observed
red shift of the H-band peak energy with time is produced by the recovery of the
space charge as the photoexcited carriers recombine. Bergman \textit{et al.}
also measured an increase in the H-band decay time for increasing photon energy and
interpreted it as coming from electrons and holes with larger and larger
separations in real space. Electrons and holes in close proximity have large
wave function overlap which gives short recombination lifetimes and larger
photon energies. Widely separated electrons and holes exhibit longer
recombination lifetimes and emit lower energy photons.

\subsection{Transient reflectivity\label{sec:tr}}

Transient reflectivity provides important information
about carrier dynamics in heterostructures.
However, as with photoluminescence studies, the interpretation of
the results is not straightforward because of  the
interaction of pump and probe beams with both the bulk buffer layer and with
the 2DEG, and because of the high spectral sensitivity of the reflectivity 
in the vicinity of the band gap and the 2DEG energy levels. 
Since the luminescence experiments show 
that electron hole segregation occurs in the first few
picoseconds, the reflectivity changes observed on a time scale of hundreds of picoseconds
to nanoseconds should correspond to the dynamics of the photoinduced carriers in
an already equilibrated, segregated charge distribution.

\begin{figure}[htb]
\begin{center}
\includegraphics[width=\figwid]{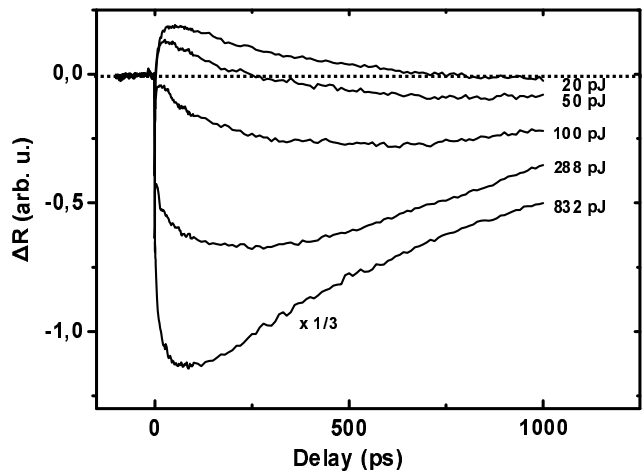}
\end{center}
\caption{Transient reflectivity $(\Delta R)$ traces on the 2DEG
sample at different energies per pump pulse. A slow decrease of
the signal in the delay range 0 to 100 ps is seen at the highest
pump power. As the pulse energy is reduced the minimum 
of the signal is reached at later times and the signal becomes
positive at early delays. The origin of this signal increase is
discussed in the main text. Data taken at 4.2 K, no magnetic
field, 780 nm pump, and weaker 820 nm probe pulses.}
\label{Transrefvsfluence}
\end{figure}
Transient reflectivity changes induced in the heterojunction
structure by the pump pulse as a function of probe delay and pump
pulse energy at 4.2 K are shown in figure~\ref{Transrefvsfluence}. The
highest excitation energy trace shows that the $\Delta R$ has an initial
slow decrease reaching a negative minimum at a delay of
approximately 100 ps. Reducing the pump pulse energy shifts this
minimum to later delays. 

Several mechanisms could be
responsible for the observed delayed formation of the $\Delta R$.
A delayed formation similar to that observed in figure~\ref{Transrefvsfluence} 
has been discussed by Prabhu \textit{et al.} \cite{Prabhu04} for $i$-GaAs. 
They showed that a delayed formation could arise at higher temperature and high excitation 
density from a combination of electron cooling, band filling and band gap renormalization.  
One could imagine that also in the present case this happens for the 2DEG and the underlying bulk GaAs, 
leading to the observed delayed formation.
The electron energy-loss, or electron cooling, in $i$-GaAs
occurs on a time scale of tens to hundred ps \cite{Alexandrou95}. 
The slow formation resulting from this should, however, speed up
as the pump pulse energy decreases \cite{Leo88}. This is in contrast to 
the observation in the HJ2DEG sample, where the
minimum of $\Delta R$ is reached at later delays when the
pump pulse energy is reduced.
Alternatively, carrier accumulation in the 2DEG by electron and hole segregation in the
built-in electric field of the buffer layer could be responsible for the 
slow formation of $\Delta R$. Again this process should slow down at
higher excitation densities. As discussed in the section on
photoluminescence (section~\ref{sec:pl}), the segregated carriers
in the buffer layer oppose the built-in electric field and reduce
it. At a low pump pulse energy the carriers are efficiently separated
by the electric field, while at higher pump pulse energy the partial
screening produced by the initial segregated carriers reduces the
built-in field, thereby slowing down the segregation of the
remaining carriers.

Even though the above discussed phenomena should play a role, 
they do not explain the observed power dependence of the 
transient reflectivity response. 
Time resolved Kerr rotation experiments (see section~\ref{sec:trkr}) 
hint toward a different interpretation. These experiments show 
the presence of at least two different photoinduced charge populations, 
each with their own $g$-factor. 
Therefore  the origin of the slow formation observed in the
heterojunction structure studied might very well be related to the
detection of carrier dynamics of two different populations, where the 
observed $g$-factors suggest that they are of 2D and 3D origin. 
Both of these carrier populations then independently produce transient
reflectivity signals that decay exponentially. 
In order to explain the observed transient response one of these 
populations should gives a positive $\Delta R$, while the other 
should give a negative $\Delta R$. Given the complex behaviour of 
the dielectric function in the vicinity of the band gap and 2DEG 
level energies, this is not unexpected. 

Based on these notions, the sum of
two mono-exponential decays of opposite sign has been 
fitted to the data of figure~\ref{Transrefvsfluence} giving 
excellent agreement. The difference between the traces at
different pump pulse energies is mainly a change in the relative
amplitude of the individual signals. The two traces taken at the
lowest pump pulse energy show a positive $\Delta R$ at early delays
that later turns into a negative $\Delta R$. The reason for this
is that the lifetime of the population giving negative $\Delta R$
is shorter than that of the other population but its amplitude is
larger. Thus, at early delays this population dominates the
signal. At later delays, on the other hand, the signal is
dominated by the population with the longest lifetime. The
estimated carrier lifetimes from fits to the data using the sum of
two exponentials with amplitudes of opposite sign are 1200 ps and
730 ps for the negative and positive contributions, respectively.
\begin{figure}[htb]
\begin{center}
\includegraphics[width=\figwid]{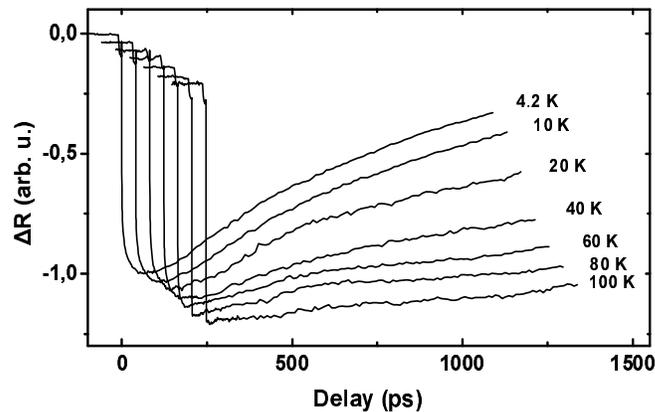}
\end{center}
\caption{Transient reflectivity of the 2DEG sample as a function of
temperature.  The lifetime of carriers giving a negative $\Delta R$ increases
with temperature in a manner comparable with confined carriers
\cite{Gurioli91}. Data taken with pump at 780 nm wavelength and pulse energy of
0.8 nJ and weaker probe at 820 nm, no magnetic field. Data offset
for clarity.} \label{TransrefvsTemp}
\end{figure}

Temperature dependent transient reflectivity measurements,
figure~\ref{TransrefvsTemp}, shed some light on the origin of the two contributions.
The observed increase of the longer lifetime
as a function of temperature is consistent with measurements by
other groups \cite{Gurioli91} of carrier recombination time constants in QWs at low
temperatures. This suggests that the negative $\Delta R$ in
figure~\ref{Transrefvsfluence} originates from photoinduced 2D electrons at the heterojunction.
The positive $\Delta R$ is then produced by 3D carriers for instance in the bulk region
of the buffer layer.

\subsection{Transient grating techniques, probing diffusion and carrier relaxation\label{sec:tg}}

One of the disadvantages of time resolved reflectivity is that for long lifetimes
diffusion of carriers out of the laser spot might play an important role.
A better, though somewhat more involved, technique to study the carrier
(and spin \cite{Weber05, Carter06}) dynamics is the so called transient grating technique.
The formation of transient gratings (TG's) is a result of a four-wave mixing
process which is described by the third order non-linear susceptibility
$\chi^{(3)}_{ijkl}$. In general $\chi^{(3)}_{ijkl}$ is a fourth rank tensor
with 81 elements. In isotropic media, however, only three elements  are
independent: $\chi^{(3)}_{xxyy}$,  $\chi^{(3)}_{xyxy}$, and
$\chi^{(3)}_{xyyx}$. $\chi^{(3)}_{xxxx}$  in this case can be expressed as:
$$\chi^{(3)}_{xxxx}=\chi^{(3)}_{xxyy}+\chi^{(3)}_{xyxy}+\chi^{(3)}_{xyyx}. $$
Indices $x$ and $y$ here refer to the polarization planes of the interacting
beams which propagate along the $z$ direction. The first and the second indices
correspond to the two pump beams which set up the holographic grating. The
third and forth indicate polarization plane of the incident probe and resulting
diffracted beam, respectively. When the polarizations of two pump beams are
parallel they interfere to form a sinusoidal intensity, and thus population
grating. In contrast, when the polarizations are perpendicular interference of
two beams is not possible, and the sample is irradiated with a uniform
intensity. Still, one a particularly interesting modulation occurs: there will
be a polarization grating in which the polarization varies sinusoidally between
left en right circular states. In magnetically active samples, this obviously
leads to a modulation of the magnetization. In other words, a spin grating will
be formed with alternating spin-up and spin-down excitations.

\begin{figure}[htb]
\begin{center}
\includegraphics[width=\figwid]{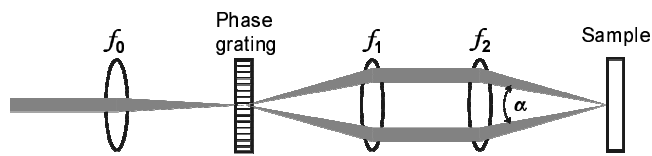} 
\end{center}
\caption{Confocal
imaging system used in the transient grating experiments, showing
the spatial arrangement of pump beams. The symbols are explained
in the main text. \label{fig:confocal}}
\end{figure}
In transient grating experiments the laser output was split by a beam splitter
into two pulses with a $3:1$ intensity ratio (pump and probe, respectively).
The pump and probe beams passed through $\lambda$/2-plates, which rotates the
plane of polarization, and subsequently through Glan Taylor polarizers to
ensure well defined polarization states. The pump beam then was focused by a
$f_0=250$~mm focal length lens onto a phase grating with a period of 30
mm$^{-1}$ which is specially designed to diffract about 30\% of the input
energy into the $+1$ and $-1$ orders of diffraction. One advantage of using a
phase grating is that it produces phase locked beams, which remain so if the same
optical elements are used for them after the grating.
In addition the use of the phase grating
ensures optimum spatial overlap of the excitation and probe pulses in the
sample\cite{Maznev98}. The two pump beams produced by the phase grating were
used to write a holographic grating on the sample. The imaging system, based on
two lenses with focal length $f_1$ and $f_2$ (see figure~\ref{fig:confocal}),
allows control of the angle between the two writing beams and so, of the
holographic grating period  $\Lambda$:
$$\alpha = 2 \arctan\left[\frac{f_1}{f_2}\left(\left(
\frac{d}{\lambda}\right)^2-1\right)^{-\frac{1}{2}}\right], \ \ \
\Lambda=\frac{\lambda}{2\sin(\alpha/2)}.$$ Here $\lambda$ is the wavelength of
the irradiation, and $d$ is the spacing of the phase grating. For room
temperature experiments $f_1$ was $250$~ mm while $f_2$ was chosen to be 150 mm
or 80 mm corresponding to grating periods of 11 $\mu$m and 5 $\mu$m,
respectively (estimated focal spot diameters: 45 and 25 $\mu$m, respectively). 
For low temperature measurements $f_2$ was replaced by a
spherical mirror with a curvature $R=-500$~mm ($f_2=250$~mm), corresponding to a grating period
of 24 $\mu$m (estimated focal spot diameter 75 $\mu$m).

The diffracted signal decays as the amplitude of the holographic grating
vanishes. The dynamics of the recorded holographic grating were probed by a
probe pulse which was delayed with respect to the pump pulses by a computer
controlled delay stage. All three, two pump and a probe, beams were arranged in
a so called BOXCAR geometry. This geometry satisfies phase matching conditions
and allows control over the direction of diffracted beam. In the BOXCAR
arrangement three parallel beams, pump $k_1$, pump $k_2$, and probe $k_3$, are
arranged parallel to each other so that they form three corners of a rectangle
in a plane perpendicular to their path. A lens (or a focusing mirror) which
center matching the one of this rectangle is used to focus the beams onto the
sample. The two pump pulses interfere in the sample and form a population
grating from which the third beam diffracts into the direction $k_s$ satisfying
the phase matching conditions, {\em i.e.} it emerges through the fourth corner
of the rectangle in the direction $k_s=k_3+k_1-k_2$. Since the diffracted
signal is typically rather weak the BOXCAR geometry is convenient because the
direction of the diffracted beam is strictly determined. During the
measurements transient reflectivity (TR) and TG signals were recorded
simultaneously by two photodiodes. This allows simultaneously determination of
both the photoinduced electron (spin) decay time, and the grating decay time in
a single scan. Since transient reflectivity gives information on the population
dynamics, and transient grating decay reflects both the spatial diffusion and the
population decay, a simultaneous measurement permits separation of these two
contributions and eventually the diffusion coefficient as detailed in 
section~\ref{sec:tg}.

A basic description of the dynamics of sinusoidal TG formed by photogenerated
excess carriers was first developed by Woerdman \cite{Woerdman71}, and reviewed
by Eichler {\em et al.}\cite{Eichler86}.
After ultrafast holographic
excitation excess carriers
simultaneously recombine and diffuse within the sample, eventually leading to
equilibrium conditions. In the two-dimensional case the interference between the two
pump pulses results in a harmonic modulation $\delta N(x,t)$ of the charge density $N(t)$:
$$\delta N(x,t)=\frac{\delta N(t)}{2}\left(1+\sin(2\pi x/\Lambda)\right),$$
where $\Lambda$ is the period of the grating.
When carrier relaxation and diffusion occurs on similar time scales,
the decay of the grating may be described by \cite{Moss81}
$$\frac{\partial \delta N(x,t)}{\partial t}=D\nabla^2(\delta N(x,t))-\frac{\delta N(x,t)}{T_r},$$
with $T_r$ the time constant for population relaxation, and $D$ the
diffusion coefficient.
If the diffusion coefficient does not depend on the space
coordinate nor on the charge density then one finds for the population modulation
$$\frac{\partial \delta N(t)}{\partial t}=
-\delta N(t)\left[\frac{4\pi^2}{\Lambda^2}D+\frac{1}{T_r}\right] ,$$
with solution
$$ \delta N(t)\propto e^{-\left[\frac{4\pi^2}{\Lambda^2}D+\frac{1}{T_r}\right]t}.$$
Since the diffracted signal is proportional to the square of population modulation
($I(t)\propto (\delta N(t))^2$ ) the diffracted signal will decay as
$$I(t)\propto e^{-2\left[\frac{4\pi^2}{\Lambda^2}D+\frac{1}{T_r}\right]t}.$$
This leads to the definition of the grating decay rate constant
$$\frac{1}{T_{gr}}= \frac{8\pi^2}{\Lambda^2}D+\frac{2}{T_r},\label{eq:tgdec}$$
which depends on the grating period $\Lambda$,
the diffusion coefficient $D$, and the population decay time constant $T_r$.
The diffusion coefficient and population decay time constant can thus be
determined independently from two grating experiments using different grating periods.
Alternatively, the diffusion constant can also be determined
from a simultaneous measurement of the dynamics of the diffracted signal and the population
relaxation (time resolved reflectivity).
Once the diffusion coefficient is known, one can calculate the mobility $\mu$ using
$$\mu=\frac{eD}{k_bT} ,$$
where $e$, $k_b$, and $T$ are the electron charge, Boltzmann coefficient and the temperature, respectively.

\begin{figure}[htb]
\begin{center}
\includegraphics[width=\tfigwid]{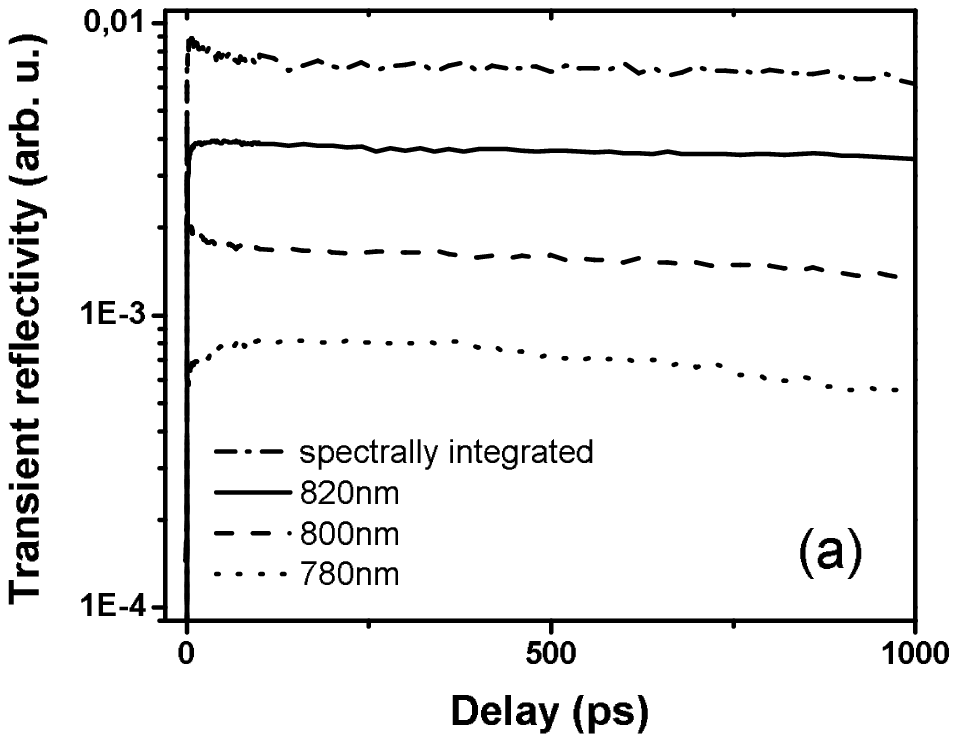}
\includegraphics[width=\tfigwid]{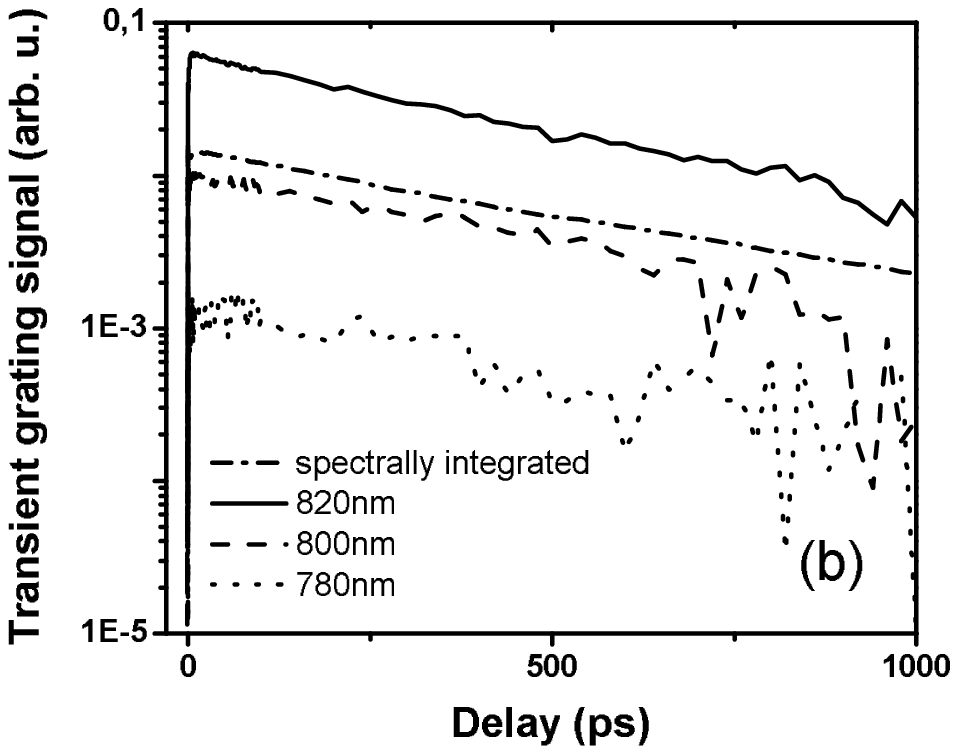}
\end{center}
\caption{The transient reflectivity (a) and transient grating (b) signals
measured on the 2DEG sample, at different wavelengths.
Data taken at 294 K, no magnetic field, 780 nm pump, probe wavelengths are
indicated in the panels. \label{fig:trgroom}}
\end{figure}
First some results obtained for a heterojunction at room
temperature are discussed, which has the advantage that one can easily change the grating
period by changing the angle of incidence for the pump beams.
Figure~\ref{fig:trgroom} shows TR and TG signals measured at room temperature
for various probe wavelengths. The pump wavelength was 780 nm
with an energy of 1125 pJ per pulse. 
Evidently, the TR dynamics is
wavelength dependent on a sub-100~ps time scale. This might result from a
variety of processes such as intraband relaxation, exciton formation, bulk
recombination and excited electron migration from the bulk to the
heterojunction as discussed in section~\ref{sec:tr}. In contrast the decay of
the TG signal is substantially less probe wavelength dependent. Apparently the
fast processes determining the initial dynamics of the TR signal do not play a
role in the TG dynamics. On the time scale on which the TG signal relaxes, the
TR dynamics are practically wavelength independent.

\begin{figure}[htb]
\begin{center}
\includegraphics[width=\figwid]{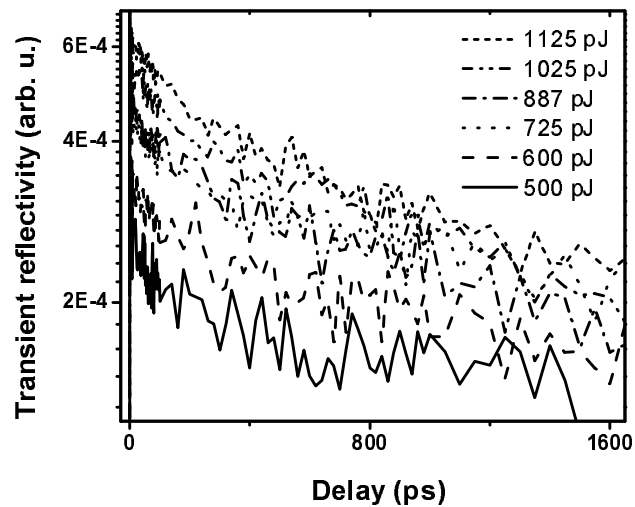} 
\end{center}
\caption{The
transient reflectivity measured on the 2DEG sample, at different
pump pulse energies. Data taken at 294 K, no magnetic field, 780
nm pump. A fraction  of the output of the Ti:Sapphire laser was
used as probe, without spectral filtering. \label{fig:trpowroom}}
\end{figure}

\begin{figure}[htb]
\begin{center}
\includegraphics[width=\tfigwid]{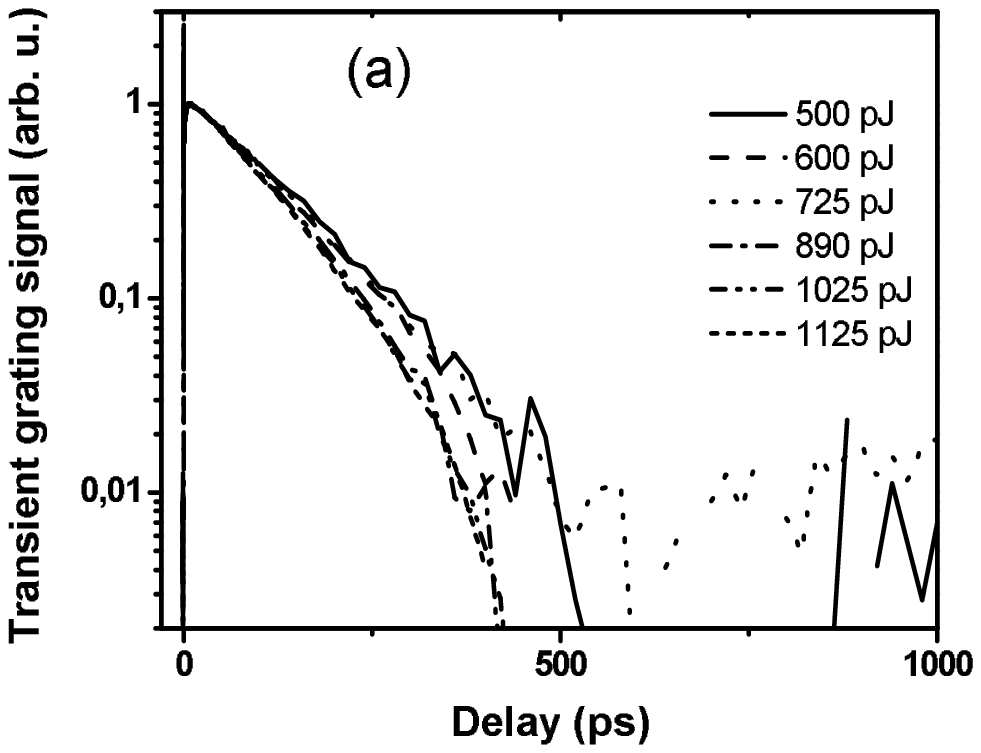}
\includegraphics[width=\tfigwid]{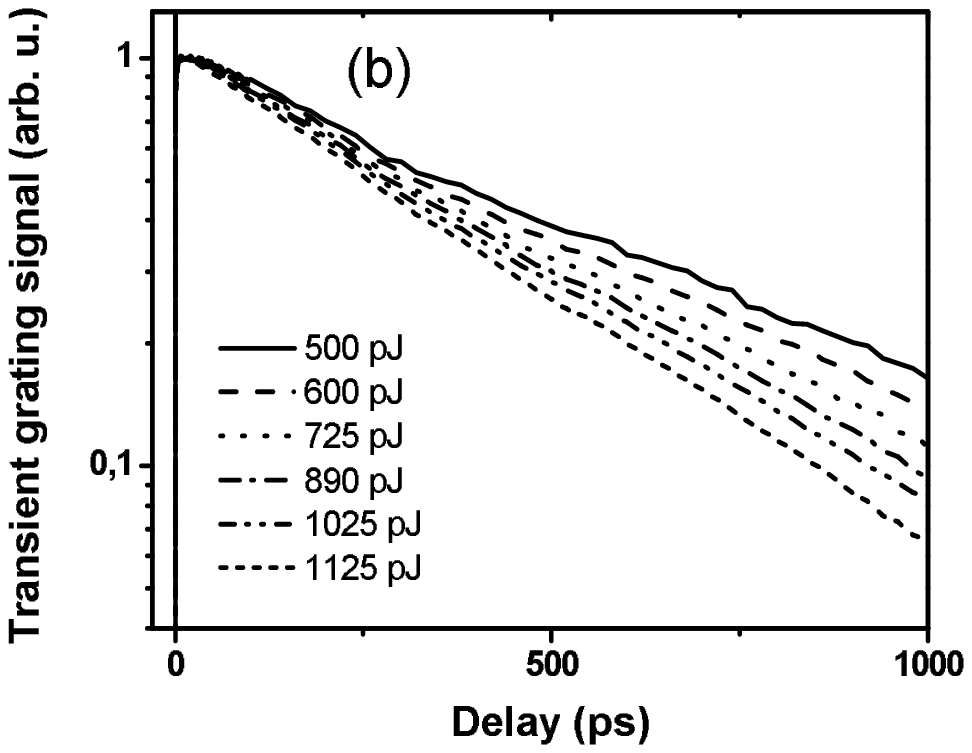}
\end{center}
\caption{Transient grating decay measured on the 2DEG sample, at
different pump pulse energies. The period of gratings is 5 $\mu$m
(a) and 11 $\mu$m (b).  Data taken at 294 K, no magnetic field,
780 nm pump. A fraction of the output of the Ti:Sapphire laser was
used as probe, without spectral filtering. \label{fig:tgpowroom}}
\end{figure}

The decay of the TR and TG signals are more strongly dependent on the excitation power,
as is shown in figures~\ref{fig:trpowroom} (TR) and \ref{fig:tgpowroom} (TG).
The TR dynamics evidently speeds up with increasing excitation power, and
is well approximated by a two-exponential decay function with typical
time constants of about 100 ps and 2 ns.
In line with this, also the TG decay, measured for grating periods 5 $\mu$m and 11 $\mu$m,
becomes faster upon increasing excitation power (see figure~\ref{fig:tgpowroom}). It
is nearly single exponential, except for a delayed formation observed during
the first tens of picoseconds (see also section~\ref{sec:tr}). 
In order to check for possible accumulation effects such as heating or
secondary excitations of long-living photoexcited species  
TR and TG measurements have been performed at different pulse repetition rates (800 kHz $-$ 4 MHz).
Since these experiments showed no dependence of the observed dynamics
on the repetition rate one may conclude that heating and secondary excitation processes do
not play an important role. 

The diffusion coefficient determined from the data of figure~\ref{fig:tgpowroom}
is tabulated in table~\ref{tab:rtdiff}. The value of about 25 cm$^2$/s, which is quite
independent on the pump power, yields a
room temperature mobility for the photoexcited carriers
of $\sim$10$^3$~cm$^2$/Vs. This is almost an
order of magnitude smaller than the value determined from room temperature
transport experiments under illumination ($8\times10^3$~cm$^2$/Vs, which is not surprising
since the TG technique is particularly sensitive to the carriers with the lowest mobility.
The complicated dynamics of the TR complicates
extraction of the diffusion constant from the simultaneous measurements of TR and
TG decay, but leads to similar values. Conversely, the
population decay constant (T$^{calc}_2$) extracted from the transient grating experiments
with two different grating spacings (5 and 11 $\mu$m)
is in reasonable agreement with the slow component observed in the TR experiments.
The independence of the diffusion constant (the electron mobility) on the excitation density shows that
the observed speeding of the decay of the TG signal is solely caused by the faster relaxation of 
electron population as also observed in the transient reflectivity data recorded under the same conditions.

\begin{table}
\centering
\begin{tabular}{ccccccc}
\hline\hline
Energy (pJ) & $T^{TR}_1$ (ps) & $T^{TR}_2$ (ns) & T$^{5 \mu\mathrm{m}}_{gr}$ (ps)& T$^{11 \mu\mathrm{m}}_{gr}$ (ps)& T$^{calc}_2$ (ns) & $D$ (cm$^{2}$/s)\\
\hline
1125 & 148$\pm$28 & 2.1$\pm$0.3& 107$\pm$1 & 365$\pm$2& 2.0 & 26.4\\
1025 & 129$\pm$25 & 2.3$\pm$0.4& 120$\pm$1 & 369$\pm$8& 1.6 & 22.4\\
 887 & 115$\pm$20 & 2.4$\pm$0.3& 124$\pm$2 & 389$\pm$3& 1.8 & 21.9\\
 725 & 108$\pm$25 & 3.0$\pm$0.9& 119$\pm$1 & 438$\pm$3& 2.9 & 24.4\\
 600 & 101$\pm$28 & 3.1$\pm$0.9& 119$\pm$1 & 478$\pm$4& 4.5 & 25.2\\
 500 &  83$\pm$20 & 3.2$\pm$0.7& 140$\pm$2 & 523$\pm$4& 3.6 & 20.9\\
\hline\hline
\end{tabular}
\caption{
The power dependent decay time constants of transient reflectivity signal ($T^{TR}_1$ and $T^{TR}_1$) and of
transient grating signals (T$^{5 \mu\mathrm{m}}_{gr}$ for the 5 $\mu$m spaced grating,
and  T$^{11 \mu\mathrm{m}}_{gr}$ for the 11 $\mu$m spaced grating). Data measured on the 2DEG sample. The last two columns
give the population decay constant (T$^{calc}_2$) and the diffusion coefficient ($D$) calculated
from the two grating time constants. Experimental parameters as in figure~\ref{fig:trpowroom}
\label{tab:rtdiff}}
\end{table}

\begin{figure}[htb]
\begin{center}
\includegraphics[width=\tfigwid]{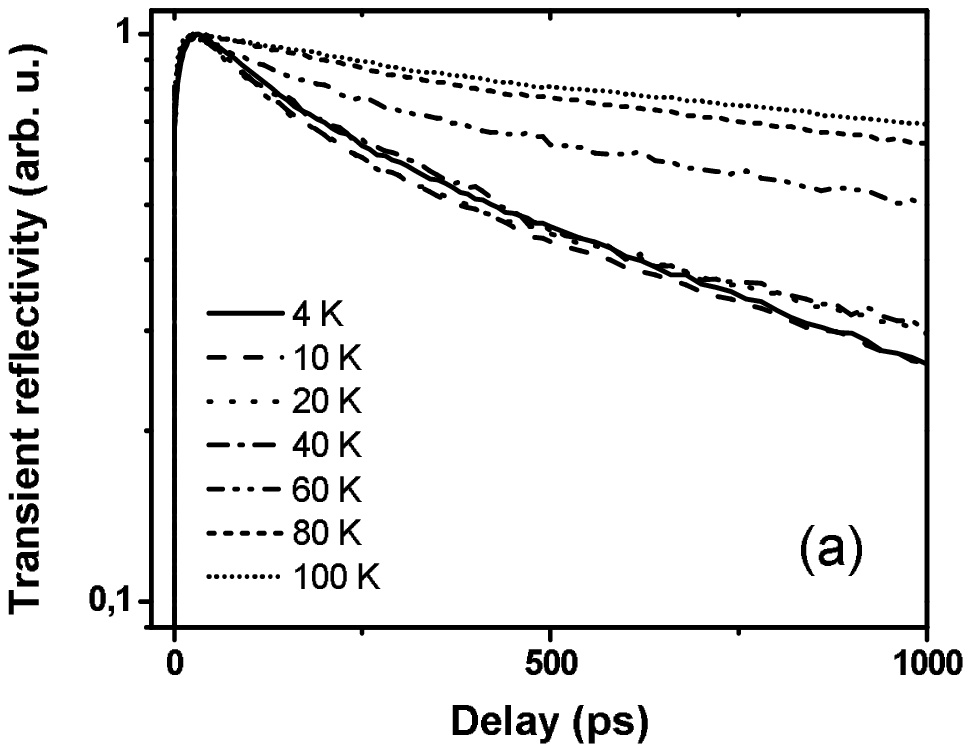}
\includegraphics[width=\tfigwid]{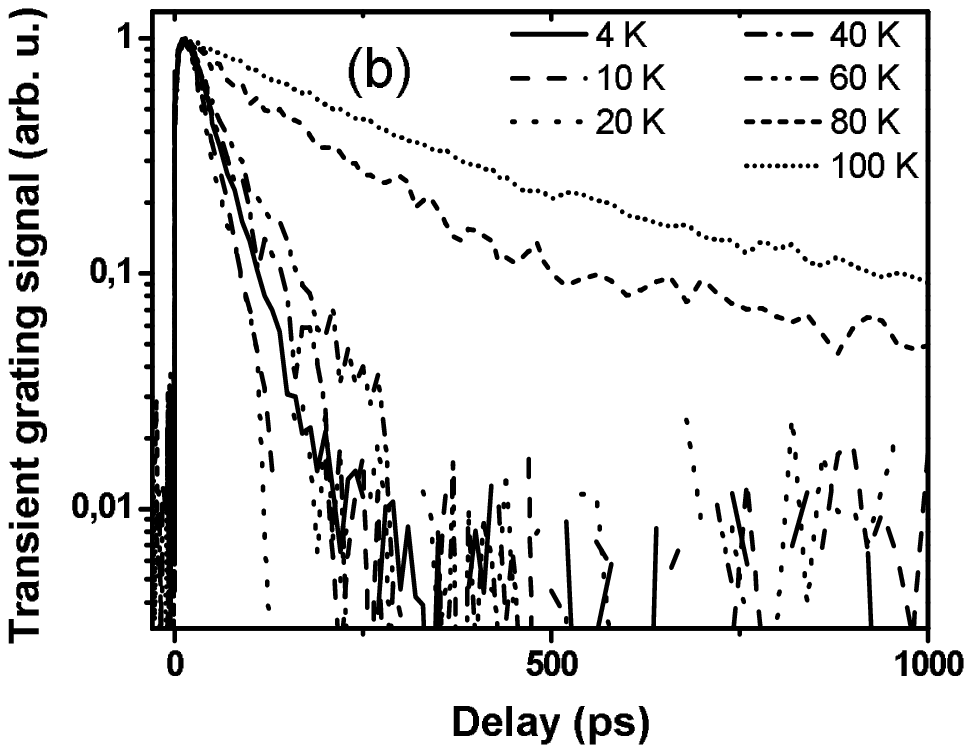}
\end{center}
\caption{The transient reflectivity (a) and transient grating (b) decay
measured on the 2DEG sample, at different temperatures.
Data taken with 780 nm pump (pulse energy 1 nJ), and weaker 820 nm
probe pulses, no magnetic field. \label{fig:tgtdep}}
\end{figure}

\begin{table}
\centering
\begin{tabular}{ccccc}
\hline\hline
Temperature (K) &  $T^{TR}$ (ns) & T$^{24 \mu\mathrm{m}}_{gr}$ (ps)
     & $D$ (cm$^{2}$/s) & $\mu$ (cm$^{2}$/Vs)\\
\hline
294$^\dagger$&-&-&26.4&$1.0\times10^{3}$\\
100&2.6&230&262&$3.0\times10^{4}$\\
 80&2.3&149&425&$6.2\times10^{4}$\\
 60&1.2&67&965&$1.9\times10^{5}$\\
 40&0.55&53&1111&$3.2\times10^{5}$\\
 20&0.36&28&2200&$1.3\times10^{6}$\\
 10&0.4&30&2077&$2.4\times10^{6}$\\
4.2&0.58&38&1667&$4.8\times10^{6}$\\
\hline\hline
\end{tabular}
\caption{
The temperature dependent decay time constants of the transient reflectivity signal ($T^{TR}$)
and of the transient grating signals (T$^{24 \mu\mathrm{m}}_{gr}$ for a 24 $\mu$m spacing grating.
The TR decay time constant was evaluated by using a single-exponential approximation in a
time window corresponding to the decay of TG signal.
The last two columns give the diffusion coefficient ($D$) and the mobility ($\mu$) calculated
from the TG and TR time constants.
Experimental parameters as in figure~\ref{fig:tgtdep}.
\label{tab:lowdiff} \\
$\dagger$ Value taken from table~\ref{tab:rtdiff}.}
\end{table}

\begin{figure}[htb]
\begin{center}
\includegraphics[width=\figwid]{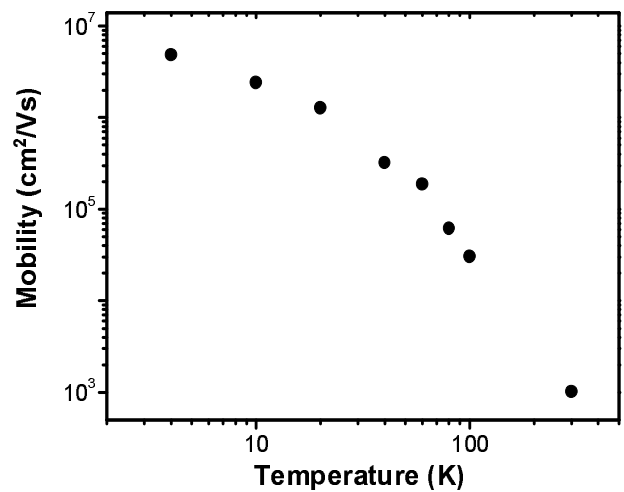} 
\end{center}
\caption{Electron
mobility of 2DEG as a function of temperature, as derived from
transient grating data as in figure~\ref{fig:tgtdep}.
\label{fig:mob}}
\end{figure}

Simultaneous temperature dependent TR and TG measurements were performed
at low temperatures using a $R=-50$~cm spherical mirror instead of the last lens in the imaging
system. This results in a grating period of 24 $\mu$m. The wavelength was 780 nm, and the excitation energy
was kept 1125 pJ per pulse throughout the experiments.
Some typical TR and TG transients measured different temperatures are shown in figure~\ref{fig:tgtdep}.
The decay of both TR and in particular the TG signals speeds up with decreasing temperature from 100 K to 4.2 K.
For the TR response this has already been discussed in section~\ref{sec:tr}.
The faster relaxation of the TG signal at low temperatures reflects a tremendous increase of
the mobility at low temperatures, typical for a heterojunction 2DEG. 
This evidently shows that the measured
signal indeed originates from the 2DEG.
The mobility as calculated from the decay of the TG and TR signals measured at
different temperatures is shown table~\ref{tab:lowdiff} and figure~\ref{fig:mob}.
At 4.2 K the mobility is in the order of $5\times10^6$ cm$^2$/Vs, which is 
comparable to the mobility determined from transport experiments 
under illumination ($2.7\times10^6$ cm$^2$/Vs), and 
substantially larger than the bulk values measured under 
similar conditions ($5\times10^4$ cm$^2$/Vs).
With increasing temperate the mobility gradually decreases
due to the increased scattering rate.
It is interesting to note that at temperatures 80 K and 100 K the TG
response becomes evidently non-single-exponential and can be reasonably
well approximated by two-exponential decay function with typical decay constants
in the range on 150 ps and a couple nanoseconds. This probably results from
the presence of two charge carrier species, one in the 2DEG, and one in the
dopant layer, with each their own temperature dependent response 
originating from the temperature dependence of the band gap.

\section{Spin dynamics\label{sec:spin}}

\subsection{Transient magneto-optics: TRKR \label{sec:trkr}}
Electron spin dynamics were investigated in the heterojunction structures by
time-resolved Kerr rotation (TRKR). This technique is sensitive to the
evolution of spin polarized carriers on ultra short time scales. A circularly
polarized pump pulse first promotes an unequal number of spin up and spin down
electrons to the conduction band of the semiconductor.
Subsequently, a
linearly polarized probe pulse is reflected from the sample.
By analyzing the change in polarization state of the probe beam
the magnetization of the sample along the direction of the probe beam 
can be traced at a given instant.\cite{Meier84}
A TRKR trace is obtained by scanning the delay between the pump and probe pulses using
a mechanical delay line.

\begin{figure}
\begin{center}
\centering\includegraphics[width=\figwid]{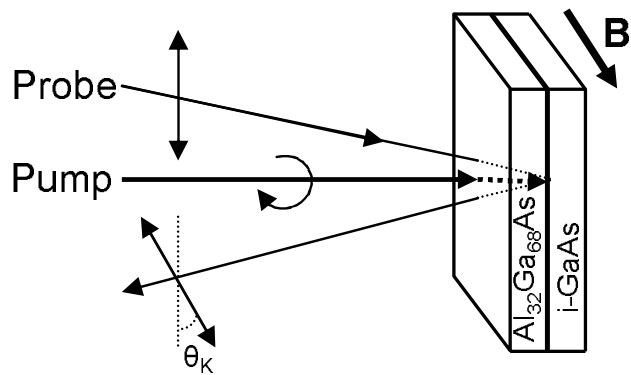}
\end{center}
\caption{Schematic of the pump and probe beams incident on the 2DEG sample as
used in TRKR experiments. The pump pulse is circularly polarized and the probe
is linearly polarized. The pulses are incident on the sample at a small angle
from the normal ($\sim$2.3$^{\circ}$). Time traces are taken by scanning the
delay between the pump and probe pulses. A rotation of the probe pulse
polarization ($\theta_{K}$) is induced by the instantaneous alignment of spins
in the conduction band of the semiconductor. The applied magnetic field is
oriented in the plane of the 2DEG.} \label{beamgeometry}
\end{figure}

The TRKR set-up is similar to the one used for the transient
reflectivity measurements, with appropriate changes for controlling the polarization state of both the pump
and probe pulses. To improve the signal to noise ratio of the signals, a lock-in technique is used with a photo-elastic modulator that alternates the polarization of the pump beam between right and
left circular polarizations at a rate of 50 kHz.
The rotation of the plane of polarization of the probe pulses,
corresponding to the Kerr rotation angle,
are detected using a balanced photodiode bridge. A schematic of the beam geometry utilized
is depicted in figure~\ref{beamgeometry}.

\begin{figure}
\begin{center}
\includegraphics[width=\lfigwid]{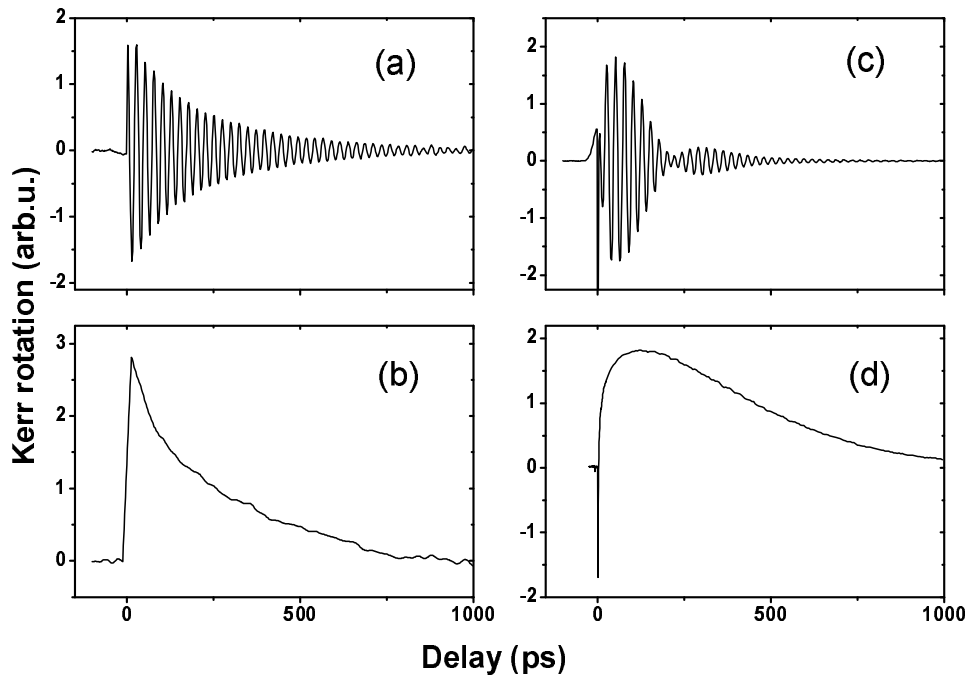}
\end{center}
\caption{Time resolved Kerr rotation signal at 4.2K from bulk $n$-GaAs
at 7 Tesla (a) and 0 Tesla (b) and from the heterojunction 2DEG also at 7 Tesla
(c) and 0 Tesla (d). The data shows considerable differences between the spin
dynamics in bulk and 2DEG samples. Most remarkable is the presence of beatings
in the signal taken at 7 Tesla in the 2DEG sample (plot c). Also in the 2DEG,
comparing plots (b) and (d) clearly shows a slow increase of the Kerr signal in
the delay range 0 to 100 ps. Data taken with 780 nm pump (0.8 nJ/pulse), 
and weaker 820 nm probe pulses. } \label{bulkand2deg}
\end{figure}

In the presence of an in-plane external magnetic field, the spins
injected by the pump pulse precess at the Larmor frequency
$\Omega_{L}$ given by:

\begin{equation}
\vec{\Omega}_{L}=g\mu_{B}\vec{B}/\hbar \label{eq:Larmorfreq}
\end{equation}

\noindent where $g$ is the electron $g$-factor, $\mu_{B}$ is Bohr's
magneton, $B$ is the applied magnetic field and $\hbar$ is the
reduced Planck constant. Spin precession shows up in TRKR traces
as distinct oscillations of the Kerr angle. 
Figure~\ref{bulkand2deg} shows low temperature TRKR traces taken on a
sample of bulk $n$-type GaAs figures~\ref{bulkand2deg}(a) and
\ref{bulkand2deg}(b) at 7 and 0 Tesla respectively. The doping
concentration of the bulk sample ($3\times10^{16}$~cm$^{-3}$ Si doping) 
was chosen equal to a
well-characterized material, that was reported to give the longest
spin coherence times in bulk GaAs \cite{Kikkawa98}. The electron
$g$-factor can be determined from the measured precession frequency
at $B\neq 0$ utilizing equation~\ref{eq:Larmorfreq}. The $g$-factor in
turn gives important information about the optically pumped spin
population. For bulk GaAs a $g$-factor of $|g|\sim 0.44$ is found,
consistent with the accepted value for this material of $g\simeq
-0.44$. The loss of coherence of the photoexcited spins can be
studied by measuring TRKR traces at $B=0$ or the envelopes of the
oscillatory signal at $B\neq 0$.

The traces taken on $n$-GaAs should be compared to the ones taken
on the heterojunction structure under similar conditions, 
figures~\ref{bulkand2deg}(c) and \ref{bulkand2deg}(d). In the HJ2DEG traces 
clear beatings are seen in the data taken at 7 Tesla, evidencing the
existence of two spin populations with different $g$-factors. The
$g$-factors determined from the data in plot~\ref{bulkand2deg}(c)
are $|g|\sim 0.44$ and $|g|\sim 0.39$. These values suggest the 
existence of both a 3D ($|g|\sim 0.44$) and a 
2D ($|g|\sim 0.39$)\cite{Hannak95} spin (and charge) population. 

The TRKR traces for the HJ2DEG sample, 
figure~\ref{bulkand2deg}(c) and (d), show a slow increase of the Kerr rotation
angle amplitude reaching a maximum at approximately 100 ps. 
The origin of this signal increase is analogous to the origin of the increase in
the transient reflectivity signal as discussed in the previous section.
Again it is needed that the two distinct carrier populations give 
Kerr rotations of opposite signs, which again may easily arise from 
the spectral details in the Kerr response. Assuming two different 
spin populations with different dephasing times and different 
$g$-factors gives a satisfactory description of the observed data.

\section{Combined spin and charge dynamics: MEOKE\label{sec:spincharge}}

\subsection{Experimental techniques for EOKE and MEOKE}

To introduce MEOKE, it is useful to first describe the electro-optical Kerr
effect (EOKE). The EOKE is the rotation of the plane
of polarization of an optical probe field in an external electric field.
Excitation of a material by an ultra short
laser pulse induces a transient anisotropy of the refractive index resulting in
a rotation of polarization and an induced ellipticity of a subsequent probe
pulse reflected of the excited material. The photoinduced anisotropy vanishes
as photoinduced excitations lose their orientational memory or relax to the
ground state. From a magneto-optical point of view EOKE  can be viewed as an
excitation of $\sigma^+$ and $\sigma^-$ transitions leading to two populations
that are coherent with the excitation. In zero magnetic field these populations
are degenerate resulting in a non-spin-polarized macroscopic dielectric
polarization $P$\cite{Worsley96}, which in turn leads to the photoinduced
anisotropy. In the presence of an external magnetic field $B$ the Zeeman
splitting of the energy levels ($\Delta E = g\mu_B B$) removes the degeneracy
of the states populated by the $\sigma^+$ and $\sigma^-$ transitions. This
leads to a macroscopic polarization $P$ which is now determined by a coherent
superposition of the energy-split states. The superposition oscillates at the
Larmor frequency $\omega=\Delta E/\hbar$ resulting in a rotation of the
polarization around the direction of $B$. The rotation of $P$ around the
magnetic field results in a modulation of the polarization of the reflected
probe beam, similar to that in the TRKR experiments. The strength of the
modulation depends on the angle at which the probe field propagates with
respect to the direction of the external magnetic field $B$. Again, as for
TRKR, for zero degrees incidence the modulation will be maximal. In general the
response will reflect both the decay of the coherence of the induced population
as well as the dephasing of the spin precession.

In the electro-optical Kerr effect (EOKE) experiments the sample was excited
with linearly polarized light while photoinduced anisotropy was recorded by
analyzing the polarization changes of a reflected weak probe beam which is
initially polarized at 45 degrees with respect to the pump. The sample's normal
was oriented at an angle of $\sim$5 degrees with respect to the propagation
direction of the probe beam. For EOKE experiments in an external magnetic field
(MEOKE), this results in a nearly transverse geometry when the applied magnetic field
is practically perpendicular to the $\mathbf{k}$-vector of the excitation and probe light,
and is parallel to the 2DEG plane of the HJ2DEG sample.

\subsection{MEOKE results}

\begin{figure}[htb]
\begin{center}
\includegraphics[width=\tfigwid]{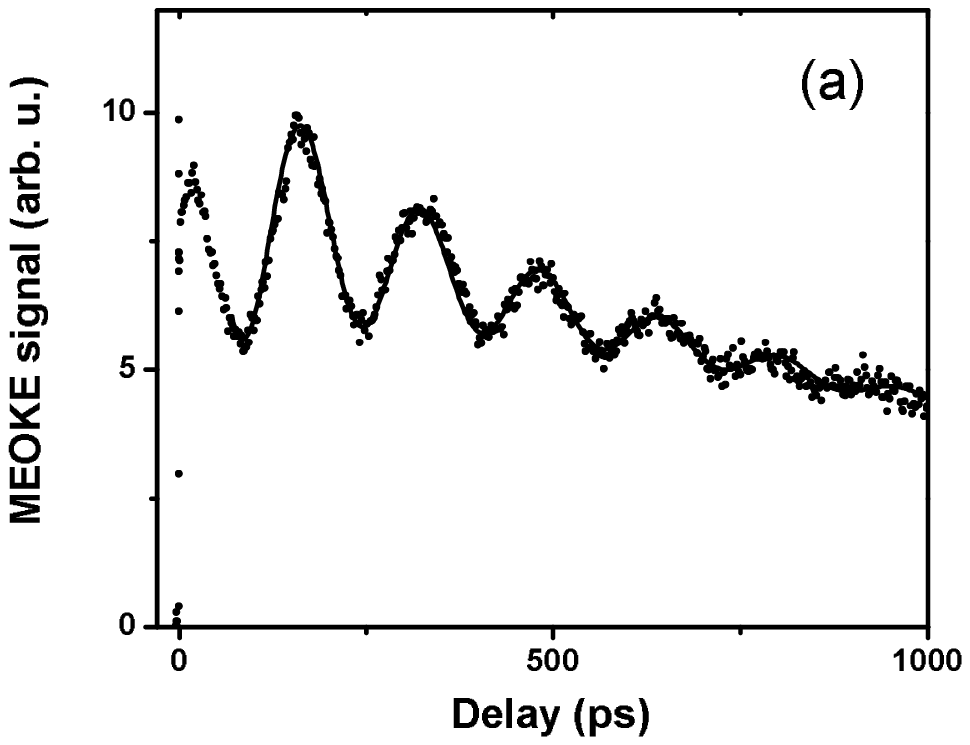}
\includegraphics[width=\tfigwid]{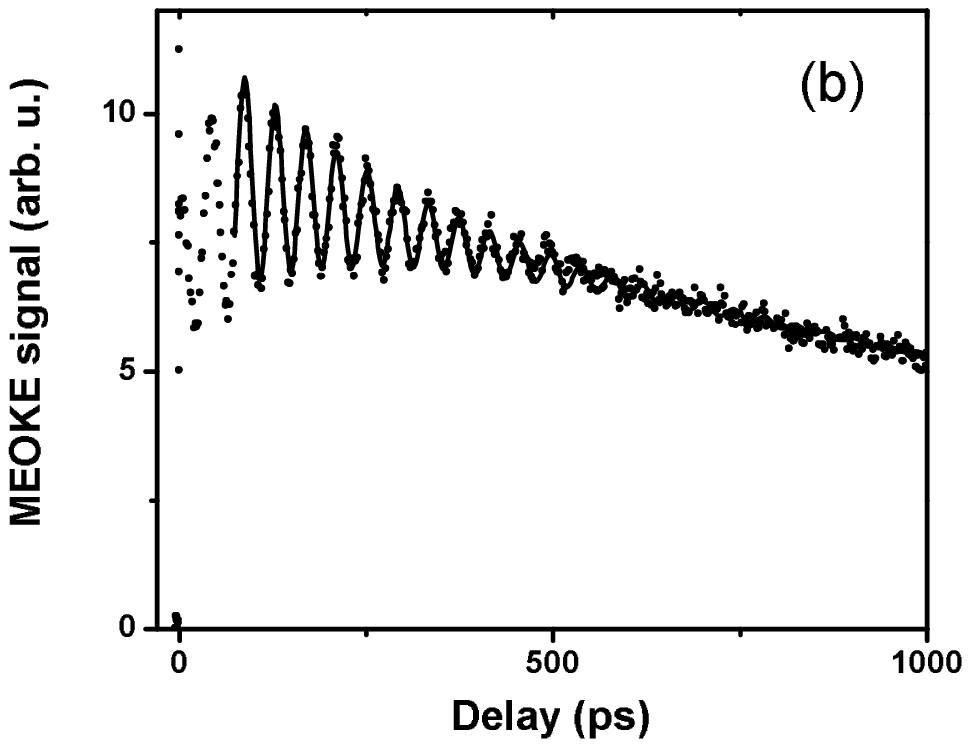}
\end{center}
\caption{MEOKE signals at 4.2 K for 1
Tesla (a) and 4 Tesla (b) applied magnetic fields, measured on the 2DEG sample.
Dots are experimental data, solid lines are fits.
Data taken with 780 nm pump (250  pJ/pulse), and weaker
820 nm probe pulses.\label{fig:meoke}}
\end{figure}

Figure~\ref{fig:meoke} shows transient birefringence decay curves at 4 K for two different
applied magnetic fields. The pump energy is 250 pJ, excitation and probe were 
centred at  780 nm and 820 nm respectively.
The time evolution of the signal can be described as a decay modulated
by oscillations. Again, an initial growth is observed, similar to the
growth seen in the other experiments. It is evident that the amplitude of the
oscillations decays substantially faster than the population decay, indicating that
the spin dephasing occurs on a shorter time scale than the loss of induced charge coherence.
Consequently MEOKE experiments potentially allow simultaneous tracing of
the dynamics of both spin and charge coherence.
\begin{table}[htb]
\centering
\begin{tabular}{cccc}
    \hline\hline
    Field (T) & $T_e$ (ps) & $T_s$ (ps) & $\omega$ (GHz)\\
    \hline
    1 & 1518$\pm$20 & 351$\pm$7 & 6.35 \\
    2 & 1487$\pm$15 & 268$\pm$5 &11.8  \\
    4 & 1444$\pm$25 & 210$\pm$9 &23.8\\
    7 & 1421$\pm$14 & 205$\pm$3 &42.3  \\
    \hline\hline
\end{tabular}
    \caption{Electron and spin life time, precession frequency at different magnetic fields from MEOKE experiments on the 2DEG sample.
    Data taken at 4.2 K, 780 nm pump (250 pJ/pulse), and weaker 820 nm probe pulses.
    \label{tab:meoke}}
\end{table}
\begin{figure}[htb]
\begin{center}
\includegraphics[width=\figwid]{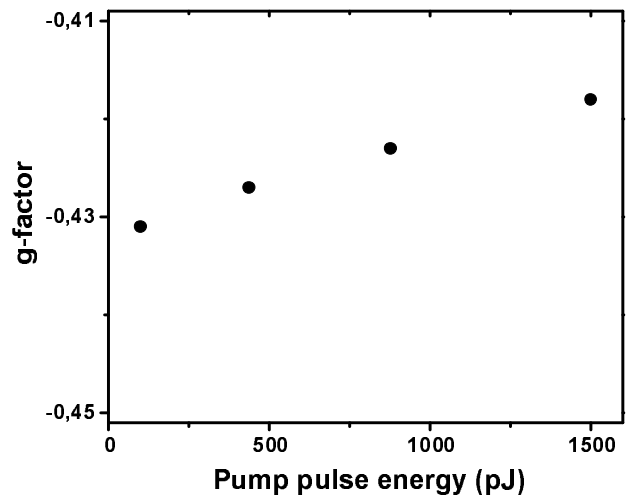} 
\end{center}
\caption{$g$-factor
derived from MEOKE measurements as a function of pump pulse energy
for the 2DEG sample, derived from data as in 
figure~\ref{fig:meoke}. The measurements were performed with 780 nm pump
and 820 nm probe pulses at temperature of 4.2 K in an external
magnetic field of 7 T. \label{fig:meokeg}}
\end{figure}
\begin{table}[htb]
\centering
\begin{tabular}{ccc}
    \hline\hline
    Energy (pJ) & $T_e$ (ps) & $T_s$ (ps) \\
    \hline
    100 & 1815 & 504  \\
    250 & 1481 & 286  \\
    437 & 1221 & 210  \\
    877 & 1040 & 118  \\
 1500 &  875 &  60  \\
    \hline\hline
\end{tabular}
    \caption{Electron and spin decay times for different excitation pulse energies,
    from MEOKE experiments on the 2DEG sample.
    The measurements were performed with 780 nm pump
    and 820 nm probe pulses at temperature of
    4.2 K in an external magnetic field of 7 T.
    \label{tab:powmeoke}}
\end{table}

\noindent
The measured traces (neglecting the initial growth)
were fitted using:
$$I(t)=A_1e^{-t/T_e}+A_2e^{-t/T_s}\sin(\omega t + \phi),$$
with $T_e$ the decay time constant of photoinduced
anisotropy, $T_s$ the spin dephasing time, and $\omega$ is the precession frequency.
The time constants and the precession frequency for different strength of
applied external magnetic field at temperature of 4 K are given in the table~\ref{tab:meoke}.
The $g$-factor deduced from these experiments is
slightly dependent on excitation density and amounts to -0.43 and -0.42 for the
excitation pulse energies of 100 pJ and 1.5 nJ respectively (see figure~\ref{fig:meokeg}).

\begin{figure}[htb]
\centering\includegraphics[width=\figwid]{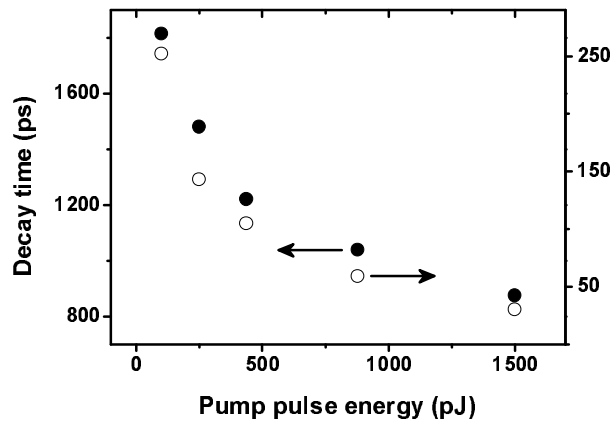} \caption{Electron
(solid circles) and spin (open circles) decay times at different
pump pulse energies for the 2DEG sample, derived from data as in 
figure~\ref{fig:meoke}.
The measurements were performed with 780 nm pump
and 820 nm probe pulses at temperature of
4.2 K in an external magnetic field of 7 T.\label{fig:powmeoke}}
\end{figure}
MEOKE experiments at different excitation densities reveal a strong
dependence of the dynamics of both charges and spins on the excitation
pulse energy, as is the case in TR and TRKR experiments. Furthermore,
the results reveal that the spin dephasing time decreases much faster than
the charge phase memory upon increasing excitation energy.
Although the nature of this behaviour is still unclear, it does show that the internal
relaxation between spin split levels strongly depends on scattering processes between
electrons in highly excited states.
Finally note that the value of the $g$-factor is slightly power dependent, resulting from the increasing
average energy of the photoinduced charge carriers. The results for several excitation
powers are summarized in figure~\ref{fig:meokeg} for the $g$-factor, and in
table~\ref{tab:powmeoke} and figure~\ref{fig:powmeoke}.

\section{Summary and conclusions\label{sec:concl}}

Optical experiments on the charge and spin dynamics in a
heterojunction structure show a quite complex behaviour. The high
mobility 2DEG does have characteristic features in for instance
the photoluminescence, the transient reflectivity, and the time
resolved Kerr rotation, but this has to be discriminated from the
features originating from in particular the 3D buffer layer. 
The interpretation of the observed phenomena is further complicated by
the changes in the HJ2DEG potential resulting from the
photoinduced charges. Nevertheless, important information can be
extracted from the optical experiments. The photoluminescence
evidences for instance the existence of indirect 2D electron $-$
3D hole exciton recombination resulting from interaction of the
2DEG charges with acceptor-bound holes in the surrounding bulk.
The transient magneto-optical Kerr rotation experiments showed the 
existence of two populations of photoexcited charge carriers. 
It appears from 
the time resolved reflectivity experiments that each of them has 
its own relaxation time and they contribute to the observed transient
reflectivity with different signs at 820 nm probe wavelength.
Size, sign and decay times of the contributions by the different
populations depend strongly not only on the probe wavelength but
also on the excitation power. This originates from the details of the
electronic structure for the 2D and 3D electrons, from carrier diffusion, 
and from optically induced band-bending and renormalization effects. 
One of the best methods to study the charge dynamics is the transient
grating technique, since it allows to discriminate between
population decay dynamics and diffusion processes. As a bonus one
also obtains the mobility of the charge carriers. A variant of
this technique, using a polarization grating rather than an
intensity grating, has also proved to be quite successful in the
study of spin dynamics \cite{Weber05,Carter06}. The most
straightforward technique to study the spin dynamics are the time
resolved Kerr rotation experiments. The two species of charge carriers 
probed in these experiments lead to a
beating of the precession oscillations in the TRKR traces. These
TRKR experiments also show that the dephasing time of the spin
population can be much faster than the charge population decay (270 ps
and 1.2 ns, respectively measured at 4.2 K and 0 Tesla with 780 nm
pump of 0.8 nJ per pulse and 820 nm probe light). Most likely,
this relatively fast dephasing results from the high mobility of
the 2DEG in combination with the Dyakonov-Perel
mechanism\cite{Dyakonov72}, in which the spin population of an
ensemble of electrons propagating in different directions dephases
due to the anisotropy of spin-orbit field. This difference in spin
dephasing and charge decay times is also found in a simultaneous
measurement of the charge and spin dynamics using the
magneto-electro-optical Kerr effect. Again the dephasing time is
found to be quite a bit faster than the charge decay. Moreover,
the spin dephasing time rapidly decreases upon increasing
excitation power. Whereas for moderate powers (100 pJ) this time
is about half a nanosecond, it is only 60 ps at high power (1500
pJ). The origin of this rapid quenching of the spin lifetime is
not quite clear presently, possibly it results from an enhanced
mobility at high excitation (through the Dyakonov-Perel mechanism)
or maybe even from enhanced spin-flip momentum scattering through
the Elliot-Yafet mechanism\cite{Elliot54,Yafet63}.

\section*{Acknowledgements}
We are grateful to Bernd Beschoten for providing us the bulk GaAs sample, and 
to Ben Hesp for help with time-resolved PL studies.
This work is supported by the MSC$^{plus}$, and by the 'Stichting voor Fundamenteel Onderzoek der Materie (FOM)',
which is financially supported by the 'Nederlandse Organisatie voor Wetenschappelijk Onderzoek (NWO)'.

\section*{References}
\bibliographystyle{iopart-num}
\bibliography{jcf}
\end{document}